\documentclass{aa}

\usepackage{graphicx}
\usepackage{longtable}
\usepackage{lscape}
\usepackage{epsfig}
\usepackage{natbib}
\bibpunct{(}{)}{;}{a}{}{,}

\begin{document}

\title{Observation of periodic variable stars towards the Galactic spiral arms by {\sc EROS II}
\thanks{This work is based on observations made with the {\sc MARLY}
 telescope of the {\sc EROS} collaboration 
at the European Southern Observatory, La Silla, Chile.}}

\author{
F.~Derue\inst{1,2}\thanks{Presently at Centre de Physique des Particules de Marseille, {\sc IN2P3-CNRS}, 163 avenue de Luminy, case 907, 13288 Marseille Cedex 09, France},
J-B.~Marquette\inst{3},
S.~Lupone\inst{3},
C.~Afonso\inst{2,4}\thanks{Presently at Department of Astronomy,  New Mexico State University, Las Cruces, NM88003-8001, U.S.A.},
C.~Alard\inst{5},
J-N.~Albert\inst{1},
A.~Amadon\inst{2},
J.~Andersen\inst{6},
R.~Ansari\inst{1}, 
\'E.~Aubourg\inst{2}, 
P.~Bareyre\inst{2,4}, 
F.~Bauer\inst{2},
J-P.~Beaulieu\inst{3},
G.~Blanc\inst{2}\thanks{Presently at University of California, Department of Physics, Berkeley, CA 97720, U.S.A},
A.~Bouquet\inst{4},
S.~Char\inst{7}\thanks{deceased},
X.~Charlot\inst{2},
F.~Couchot\inst{1}, 
C.~Coutures\inst{2}, 
R.~Ferlet\inst{3},
P.~Fouqu\'e\inst{8,9}
J-F.~Glicenstein\inst{2},
B.~Goldman\inst{2,4}{$^{\star\star\star}$},
A.~Gould\inst{4,10},
D.~Graff\inst{2,11},
M.~Gros\inst{2}, 
J.~Ha\"{\i}ssinski\inst{1}, 
J-C.~Hamilton\inst{4}\thanks{Presently at ISN, {\sc IN2P3-CNRS}-Universit\'e Joseph-Fourier, 53 avenue des Martyrs, 38026 Grenoble Cedex, France},
D.~Hardin\inst{2}\thanks{Presently at LPNHE, {\sc IN2P3-CNRS}-Universit\'es Paris VI et VII, 4 place Jussieu, F-75252 Paris Cedex 05, France},
J.~de Kat\inst{2},
A.~Kim\inst{4}{$^{\dagger}$},
T.~Lasserre\inst{2},
L.~Le Guillou\inst{2},
\'E.~Lesquoy\inst{2,3},
C.~Loup\inst{3},
C.~Magneville \inst{2}, 
B.~Mansoux\inst{1}, 
\'E.~Maurice\inst{12}, 
A.~Milsztajn \inst{2},  
M.~Moniez\inst{1},
N.~Palanque-Delabrouille\inst{2},
O.~Perdereau\inst{1},
L.~Pr\'evot\inst{12},
N.~Regnault\inst{1}{$^{\dagger}$},
J.~Rich\inst{2}, 
M.~Spiro\inst{2},
A.~Vidal-Madjar\inst{3},
L.~Vigroux\inst{2},
S.~Zylberajch\inst{2}
\\   \indent   \indent
The {\sc EROS} collaboration\\
}

\institute{
Laboratoire de l'Acc\'{e}l\'{e}rateur Lin\'{e}aire,
{\sc IN2P3-CNRS}-Universit\'e de Paris-Sud, B.P. 34, 91898 Orsay Cedex, France
\and
{\sc CEA}, {\sc DSM}, {\sc DAPNIA},
Centre d'\'Etudes de Saclay, 91191 Gif-sur-Yvette Cedex, France
\and
Institut d'Astrophysique de Paris, {\sc INSU-CNRS},
98~bis Boulevard Arago, 75014 Paris, France
\and
Coll\`ege de France, {\sc LPCC}, {\sc IN2P3-CNRS}, 
11 place Marcelin Berthelot, 75231 Paris Cedex, France
\and
{\sc DASGAL}, {\sc INSU-CNRS}, 77 avenue de l'Observatoire, 75014 Paris, France
\and
Astronomical Observatory, Copenhagen University, Juliane Maries Vej 30, 
2100 Copenhagen, Denmark
\and
Universidad de la Serena, Facultad de Ciencias, Departamento de Fisica,
Casilla 554, La Serena, Chile
\and
Observatoire de Paris, {\sc DESPA}, 92195 Meudon Cedex, France
\and
European Southern Observatory (ESO), Casilla 19001, Santiago 19, Chile
\and
Ohio State University, Department of Astronomy, Columbus, OH 43210, U.S.A.
\and
University of Michigan, Department of Astronomy, Ann Arbor, MI48109, U.S.A.
\and
Observatoire de Marseille, {\sc INSU-CNRS},
2 place Le Verrier, 13248 Marseille Cedex 04, France
}

\offprints{
J-B. Marquette~: marquett@iap.fr
}

\date{
Received $<$date$>$ / Accepted $<$date$>$
}


\abstract{
We present the results of a massive variability search
based on a photometric survey of a six square degree region along the Galactic 
plane at ($l = 305^\circ$, $b = -0.8^\circ$) 
and ($l = 330^\circ$, $b = -2.5^\circ$).
This survey was performed in the framework of the {\sc EROS II} 
(Exp\'erience de Recherche d'Objets Sombres) microlensing program.
The variable stars were found among 1,913,576 stars
that were monitored between April and June 1998 in two passbands, 
with an average of 60 measurements.
A new period-search technique is proposed which makes use of
a statistical variable that characterizes the overall regularity
of the flux versus phase diagram.
This method is well suited when the photometric data are 
unevenly distributed in time, as is our case. 
1,362 objects whose luminosity varies were selected.
Among them we identified 9 Cepheids, 19 RR Lyr{\ae}, 34 Miras,
176 eclipsing binaries and 266 Semi-Regular stars.
Most of them are newly identified objects.
The cross-identification with known catalogues 
has been performed.
The mean distance of the RR Lyr{\ae} is estimated to be 
$\sim 4.9 \pm 0.3$ {\rm kpc} undergoing an average absorption 
of $\sim 3.4 \pm 0.2$ magnitudes. 
This distance is in good agreement with the one of disc stars which contribute 
to the microlensing source star population.
Our catalogue and light curves are available electronically 
from the {\sc CDS}, Strasbourg and from our Web site\footnote{http://eros.in2p3.fr}.
}
\titlerunning{Observation of periodic variable stars}

\authorrunning{The {\sc EROS} collaboration}

\maketitle

\section{Introduction}\label{introduction}
In 1996, the {\sc EROS II} collaboration started an
observation program towards the Galactic Spiral Arms (GSA) 
dedicated to microlensing events.
Since then, four regions of the Galactic plane located at large angles 
with respect to the 
Galactic Centre are being monitored to disentangle the disc, bar 
and halo contributions to the microlensing optical depth. 
Seven microlensing event candidates have already been published, based on 
three years (1996-98) of observations 
\citep{ErosGSA2, ErosGSA3}[hereafter papers I and II].
The distance of the source stars used in these papers 
to compute the expected optical depths 
was deduced from a detailed study of our colour-magnitude diagrams.
It was thus found that the source star population is located 
$\sim$ 7 {\rm kpc} away,
undergoing an interstellar extinction $A(V)$ of about 3 magnitudes
(see \citet{Mansoux} for more details).
This distance estimate is in rough agreement with the distance 
to the spiral arms obtained by \citet{geo94} and \citet{rus98},
but its uncertainty is limiting 
further interpretation of our microlensing optical depth estimates.
It was therefore desirable to seek more information on the
distance distribution of the source stars -- whether these
stars belong to the disc or to the spiral arms -- and on the
reddening along our observation line of sights.
This led us to perform a dedicated variable star search
 between April and June 1998, 
on a subset of our Galactic plane fields.
The analysis was restricted to the brightest 
stars of this subset.

Among the wide variety of variable stars, periodic ones 
are of particular interest. 
The properties of Cepheids make them well suited 
to trace the Galactic spiral arms. 
Their reddening is measurable as well as 
their distance via the period-luminosity (PL) relation.
RR Lyr{\ae} stars are old stars, 
well suited to trace the disc population. 
One can infer their mean dereddened magnitude 
and their absolute magnitude \citep{gould-1998}.
The infrared PL(K) relation for Miras 
and Semi-Regular variable stars can be 
calibrated using a comparison of {\sc DENIS} and 
{\sc EROS} {\sc LMC} giant stars \citep{Cioni-2001}.
Finally detached eclipsing binaries also offer the opportunity to measure 
their stellar parameters and their distance \citep{paczynski-1996}. 

This paper presents the results of this particular campaign which led to 
a catalogue containing a large number of new 
variable objects in the Galactic plane.
Sect. \ref{Exp} gives the basic features of the observational setup, 
Sect. \ref{Psearch} gives details on a new algorithm used 
to search for periodic variations of the luminosity.
Sect. \ref{catalog} describes the catalogue and the cross-identification 
process.
In Sect. \ref{discussion} we use the selected RR Lyr{\ae} to 
estimate the mean reddening of our fields and we give the distance 
distribution of these stars.
\section{Experimental setup and observations \label{Exp}}
The {\sc MARLY} telescope and its two cameras, the way we carry 
out our observations, as well as our data reduction sequence 
are described in Paper I and references therein.
The two {\sc EROS} passbands are non standard.
The so-called {\sc EROS}-red passband $R_{E}$ is centred on 
$\bar\lambda = 762\ {\rm nm}$, close to $I_C$ Cousins,
with a full width at half maximum $\Delta\lambda \simeq 85\ {\rm nm}$,
and {\sc EROS}-visible passband $V_{E}$ is centred on 
$\bar\lambda = 600\ {\rm nm}$, close to $V_J$ Johnson,
with $\Delta\lambda \simeq 78\ {\rm nm}$.  
The {\sc EROS II} colour magnitude system is defined as follows~:
a zero colour star with $V_E - R_E = 0$ (a main sequence A0 star) will 
have its $R_E$ magnitude numerically equal to its Cousins $R_C$ magnitude 
and its $V_E$ magnitude numerically equal to its Johnson $V_J$ 
magnitude.
The colour transformation between the {\sc EROS II} system ($V_E$,$R_E$) 
and the standard Johnson-Cousins ($V_J$,$I_C$) system is then~:
\begin{eqnarray}
I_C &=& R_{E} + 0.01 \times (V_{E} - R_{E}) \label{colour} \\
V_J &=& V_{E} + 0.39 \times (V_{E} - R_{E}) \ . \nonumber
\end{eqnarray}
The colour coefficients are obtained from the study of our passbands 
based on Landolt standards and on one of the {\sc EROS II LMC} 
fields observed simultaneously 
in $B$$V_J$$R_C$$I_C$ with the Danish 1.54 {\rm m} 
(at {\sc ESO}-La Silla) and 
with the {\sc MARLY}.
The zero points are established with tertiary standards 
in $BVRI$ taken with the Danish 1.54 \citep{Regnault}.
We have cross-checked our $R_E$ photometry with the 
 $I_C$ one of {\sc DENIS} \citep{Fouque}.
Furthermore, using Eqs.~(1),
the mean magnitudes of the {\sc LMC} red-giant clump stars agree 
within 0.1 magnitude with determinations made by \citet{harris99} and 
\citet{ogle98}.
We thus estimate that the precision of the zero points of the 
{\sc MARLY} calibration is $\sim 0.1^{\rm mag}$.
%
\begin{table}[h!]
\begin{center}
\caption{ Characteristics of the 6 one square degree
fields monitored for this study. 
The equatorial (J2000) and galactic coordinates of the field centres, 
the number of photometric measurements per light curve per colour $N_{m}$
and the number of analysed light curves $N_{obs}$ (in millions) are given 
for each field. 
The observation duration was $\sim$ 100 days.
} 
\label{tabfields}
{
\begin{tabular}{|c|c|c|c|c|c|c|}\hline
\multicolumn{1}{|c}{Field}&
\multicolumn{1}{|c|}{$\alpha$}&
\multicolumn{1}{|c|}{$\delta$}&
$b^{\circ}$ & $l^{\circ}$ & $N_{m}$  & $N_{obs}$  \\ \hline
\multicolumn{5}{|c|} {Norma ($\gamma$ Nor)}  & 59  & 1.30 \\ \hline 
                   
gn450 &  16:09:45 & -53:07:03  & -1.17 & 330.49 & 63  &0.28\\
gn453 &  16:22:28 & -52:06:20  & -1.69 & 332.24 & 52  &0.36\\
gn455 &  16:26:52 & -52:21:02  & -2.35 & 332.54 & 57  &0.30\\
gn459 &  16:15:51 & -54:48:45  & -2.86 & 329.82 & 61  &0.36 \\      \hline
\multicolumn{5}{|c|} { Musca ($\theta$ Mus)} &   60 & 0.61 \\ \hline  
               
tm550 &  13:27:04 & -63:02:18 & -0.47 & 306.98 & 61  &0.27 \\          
tm551 &  13:31:18 & -63:34:41 & -1.07 & 307.37 & 60  &0.34 \\ \hline      
\multicolumn{6}{|r|} { Total} & 1.91 \\ \hline  
  
\end{tabular}
}
\end{center}
\end{table}

Among the 29 fields of the {\sc EROS} GSA microlensing program, 
the six fields considered here were monitored about once per 
night between April and June 1998.
They represent $2$ square degrees towards $\theta$ Mus and 
$4$ towards $\gamma$ Nor (named after the closest bright star).
Table \ref{tabfields} gives their coordinates, 
the number of images $N_m$ taken in each direction and the 
number of analysed light curves $N_{obs}$.
\begin{figure}[h!]
\begin{center}
\resizebox{\hsize}{!}{\includegraphics{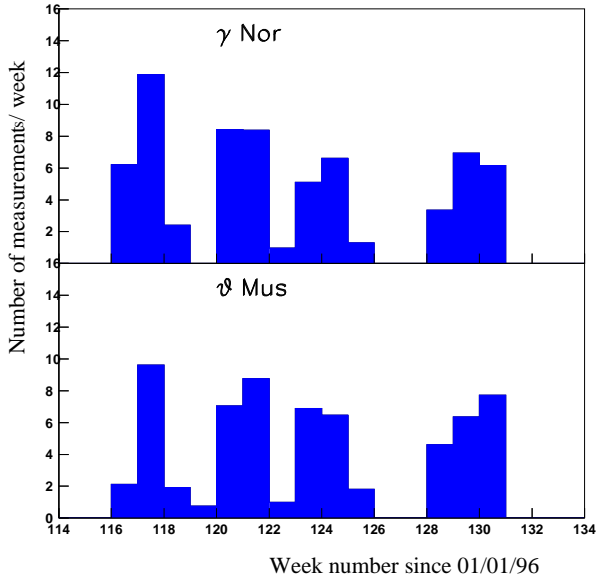}}
\caption[]{Average time sampling for the 6 fields monitored
(2 towards $\gamma$ Nor, upper panel, and 4 towards $\theta$ Mus, 
lower panel), in number of measurements per week and per field. }
\end{center}
\label{sampling}
\end{figure}
To avoid CCD saturation by the brightest stars ($I_C \sim 9$), 
by Cepheids or Miras in particular, 
we have reduced the exposure time to 15 {\rm s} instead of 
the 120 {\rm s} used in the microlensing survey.
As a consequence the catalogue is incomplete as far as faint stars 
are concerned, but it could be 
updated later by using the total set of available GSA images.
Fig.~1 shows the average time sampling.
Three gaps can be seen in our data~: the first two (around 
weeks 119 and 123) were due to bad weather conditions while the 
third one (around week 127) corresponds to the 
annual maintenance of our setup.
\section{Search for periodic stars \label{Psearch}}
\subsection{Reconstruction of the light curves}
Since the {\sc EROS} photometry is described in detail in \citet{PEIDA}
only the main features of the {\sc PEIDA++} package are summarised below.
For each field, a template image is first constructed using 
one exposure of very good quality. 
A reference star catalogue is set up with this template 
using the {\sc CORRFIND} star finding algorithm \citep{Eros2SMC1}. 
For each subsequent image, after geometrical alignment with the template, 
each identified star is fitted together with 
its neighbours, using a {\sc PSF} determined on bright isolated 
stars and imposing the position from the reference catalogue. 
A relative photometric alignment is then performed, assuming that 
most stars do not vary. 
Photometric errors are computed for each measurement, assuming again 
that most stars are stable, and parametrised as a function of star 
brightness and image sequence number. 
Fig.~2 shows the mean point-to-point relative dispersion 
of the measured fluxes along the light curves as a function of $R_{E}$ 
and $V_{E}$. 
The photometric accuracy is $\sim$15\% at $R_{E}$$\sim$18, 
and about 2\% for the brightest stars. 
\begin{figure}[h!]
\begin{center}
\resizebox{\hsize}{!}{\includegraphics{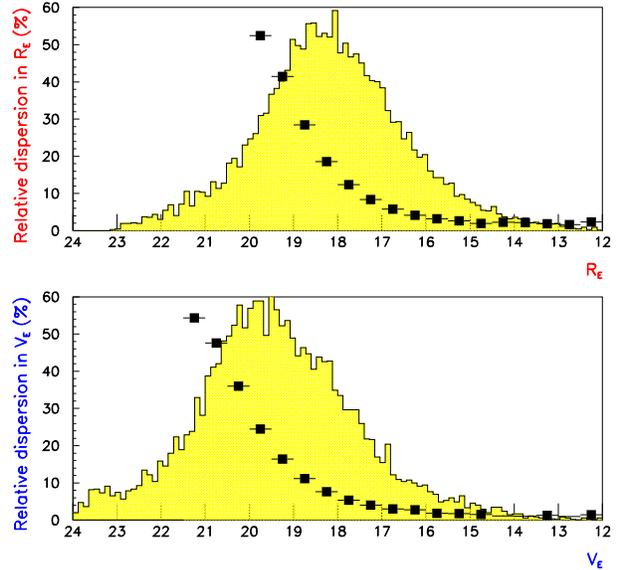}}
\caption[]{Average value of the relative frame to frame dispersion 
of the luminosity
measurements versus $R_{E}$ (upper panel) and $V_{E}$  
(lower panel), for stars with at least 30 reliable
measurements in each colour.
This dispersion is taken as an estimator of the mean photometric precision.
The superimposed hatched histograms show the {\sc EROS} 
magnitude distribution of monitored stars.}
\end{center}
\label{resolution}
\end{figure}
Finally, using the {\sc PEIDA++} photometric package, 
we reconstruct the light curves of 1,913,576 stars.
%
\begin{table}[h!]
\caption[]
{Impact of each selection criterion on the data. For each cut, 
the number of remaining light curves is given.}
\label{tabselection}
\begin{flushleft}
\begin{center}
\begin{tabular}{llr}
\hline
Cut &         Criterion                       \\ \hline
& Total analysed                            & 1,913,576    \\
1 & $N_{m}$ $>$ 30                          & 1,299,690  \\
2 & $R_{E} < 17$                            & 330,089     \\
3 & Pre-filtering                           & 41,545     \\
4 & Period search                           & 2,553      \\
5 & Aliasing                                & 2,424      \\
6 & Visual inspection                       & 1,362      \\ 
\hline
\end{tabular}
\end{center}
\end{flushleft}
\end{table}
\subsection{Pre-selection}
Each one of the light curves is subjected to a series of selection criteria 
in order to isolate 
a small sub-sample on which we will apply the time consuming period search 
algorithms.
These analysis cuts are briefly described hereafter 
(see \citet{TheseDerue} for more details) and their 
effect on the data is summarised in Table \ref{tabselection}~:
\begin{itemize}
\item[{\it cut 1}~:] At least 30 measurements should be available 
in both passbands 
and the base flux must be positive;
\item[{\it cut 2}~:] The search is restricted to stars whose magnitude is 
$R_{E}$$<$$17$ which corresponds to a photometric 
accuracy in $R_{E}$ better than $\sim$10\%.
\item[{\it cut 3}~:] A non specific pre-filter is applied which retains most  
variable stars. It selects light curves satisfying one or both 
of the following criteria~:
\begin{itemize}
\item the relative dispersion of the flux measurements is 
25\% larger than the average one for the set of stars having 
the same magnitude;
\item the distribution of the deviations with respect to the 
base flux is incompatible with the one expected from a stable source 
with Gaussian errors during the observation period (Kolmogorov-Smirnov test). 
\end{itemize}
These cuts are tuned to select $\sim$10\% of the light curves. 
We have checked that this procedure allows one to retrieve the previously 
known Cepheids observed by {\sc EROS} in the Magellanic Clouds. 
We also keep a randomly selected set of light curves ($\sim$2\%) 
to produce unbiased colour-magnitude diagrams, for comparison purposes;
\end{itemize}
At this stage a set of 41,545 light curves remains 
which is then subjected to a periodicity search.
\subsection{Light curve selection}
We use three independent methods to extract periodic light curves.
The first two are classical methods already described in the literature~: 
method 1 is based on the Lomb-Scargle periodogram \citep[see][]{Scargle} 
while method 2 makes use of the One Way Analysis of Variance algorithm 
\citep[see][]{Czerny}.
Both provide the probability for false periodicity detection.
In method 1 one computes the Fourier power over a
set of frequencies. 
It is therefore well adapted to identify sinusoidal light curves. 
It can be improved by incorporating 
higher harmonics in order to detect 
any kind of variability such as eclipsing 
binaries \citep{grison, grison-1995}.

We developed a new method, the third one, in order to extract periodic 
light curves in a way which is insensitive to the particular shape 
of the variation.
This method also provides a probability for false periodicity detection. 

\begin{figure}[h!]
\begin{center}
\resizebox{\hsize}{!}{\includegraphics{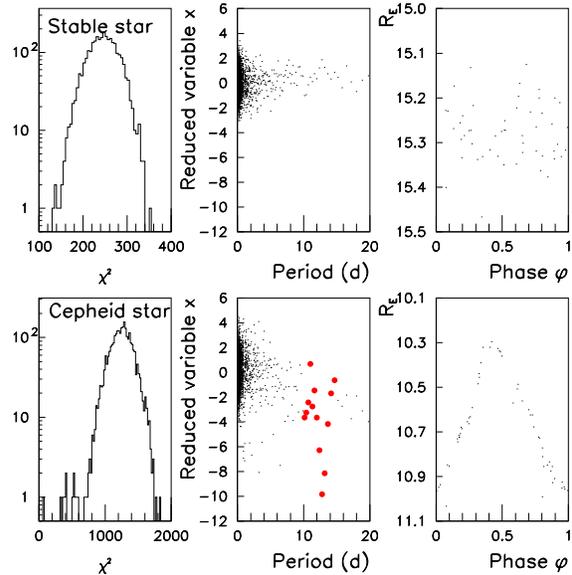}}
\caption[]{Distributions of $\chi^2$ (left panels) and
variation of $x$ with the value of the test-period in days (middle panels) 
obtained with the third method 
for a stable star and for a typical Cepheid candidate. 
The bold dots are pointing to the actual period of the star. 
The right panels show the light curve (in $R_E$) obtained once 
the period has been folded in.
}
\end{center}
\label{periodogram}
\end{figure}
It consists in searching for a frequency such that the corresponding 
phase diagram, {\it i.e} the series of fluxes $F_i$ 
versus phases $\varphi_i$ in increasing order of $\varphi$,
displays a regular structure significantly less scattered 
than for other frequencies.
Let $T_{obs}$ be the observation duration ($\sim 100$ days in this analysis). 
We span the frequency domain from $T_{obs}^{-1}$ to 
$(0.2\ {\rm days})^{-1}$ with 
a constant step of $(4 \times T_{obs})^{-1}$. 
The total number of test-frequencies is thus $N_{test}~\sim~2000$.
This sampling ensures that the total phase increment over $T_{obs}$
is $\pi/2$ for two adjacent test-frequencies.

For each value of the test-frequency we compute the corresponding 
phase diagram.
We calculate a $\chi^2$ from the weighted differences of $F_i$ and
the fluxes interpolated between $F_{i-1}$ and $F_{i+1}$~:
\begin{eqnarray}
\chi^2 &=& \sum_{i=1}^{N_{m}}\left(\frac{F_i-(1-R_i)\times F_{i-1}-R_i\times F_{i+1}}{\sigma_i}\right)^2 \label{chi_2}
\end{eqnarray}
where $R_i = (\varphi_i - \varphi_{i-1})/(\varphi_{i+1}-\varphi_{i-1})$ 
and $N_{m}$ is the number of measurements.
The uncertainty $\sigma_i$ takes into account the errors $\epsilon_i$ on 
the flux $F_i$ and on the interpolated flux~:
$\sigma_i^2 = \epsilon_i^2 + (1-R_i)^2 \epsilon_{i-1}^2 + R_i^2 \epsilon_{i+1}^2$.
\begin{figure}[h!]
\begin{center}
\resizebox{\hsize}{!}{\includegraphics{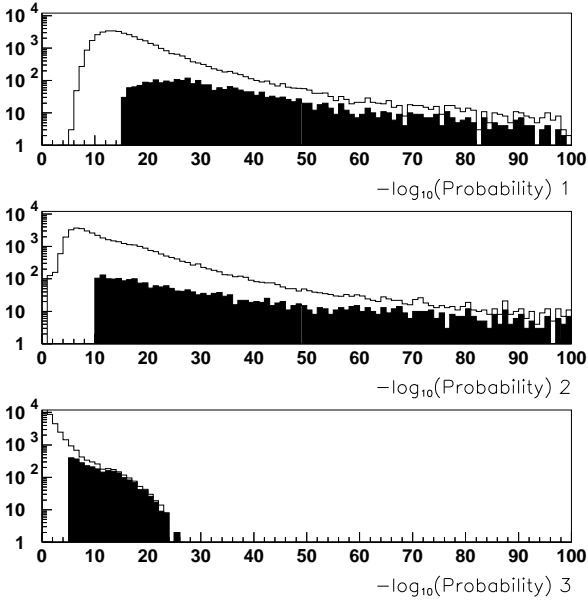}}
\caption[]
{Distribution of $-\log(Prob)$ for method 1 (top), 
2 (middle) and 3 (bottom). 
The histograms show the distribution for all 41,545 
stars on which the periodicity search is done (in white) and 
for the 2,553 selected light curves (in black).}
\end{center}
\label{distribproba}
\end{figure}
Expression (\ref{chi_2}) can be interpreted as the $\chi^2$ of the 
set of differences between the odd measurements with respect 
to the line joining even ones, added to the $\chi^2$ of the 
set of differences between even measurements with respect to the 
line joining odd ones.
If the star is measured in both colours, 
we add the $\chi^2$ obtained in each colour ($N_{m}$ is then twice 
as large).  
For a given stable star with Gaussian errors,
each phase diagram can be considered as a random realisation 
of the light curve.
When the test-frequency spans the search domain, the distribution
of the $\chi^2$ parameter defined by Eq.~(2) is the one of the 
standard $\chi^2$ with $N_{m}$ degrees of freedom.
Since $N_{m}$ is large enough, this distribution is close to a 
Gaussian with average $N_{m}$ and variance $2N_{m}$ 
(see upper left panel of Fig.~3).
For a periodic variable star,
the $\chi^2$ distribution displays a main cluster, 
when the test-frequency results in a phase diagram with non-correlated 
point to point variations, and a few lower values
when the test-frequency corresponds to a phase diagram with a regular 
structure (see lower left and middle panels of Fig.~3).
In practice, instead of using the parameter defined by 
Eq.~(\ref{chi_2}), we use the reduced variable~:
\begin{eqnarray}
x &=& \frac{\chi^2-<\! \chi^2 \!>}{<\! \chi^2 \! >/N_{m}}
\times\frac{1}{\sqrt{2N_{m}}} \label{x}
\end{eqnarray}
where $<\!\chi^2\!>$ is the average of the realisations of $\chi^2$
for all test-frequencies.
For a stable star, the distribution of this variable $x$ is a Gaussian 
centred at zero, with unit variance.
If the errors are correctly determined $<\! \chi^2 \! >/N_{m}$ is 
close to unity;
if the errors are all systematically overestimated (or underestimated), 
then including this term ensures a global renormalisation 
of the errors in Eq.~(\ref{x}), and the distribution of our reduced variable 
$x$ is also a normal distribution.
Let $x_{min}$ be the smallest value of $x$ calculated among 
all test-frequencies for a given star.
Under the hypothesis that the light curve is produced by a stable star,
the probability to obtain at least
one value $x \leq x_{min}$ in a series of $N_{test}$ realisations is
$Prob(x < x_{min} | N_{test}) = 1 - \left[ 1 - \frac{1}{\sqrt{2\pi}} \int_{-\infty}^{x_{min}} e^{-x^{2}/2} d\!x \right]^{N_{test}}$.
If this probability is small, then~:
\begin{eqnarray}
Prob(x < x_{min} | N_{test}) &\simeq& \frac{N_{test}}{2} {\rm erfc} 
\left( \frac{- x_{min}}{\sqrt{2}} \right) \ . \label{P_small}
\end{eqnarray}
If the light curve exhibits periodic variations, then there exist
test-frequencies for which $x$ is significantly smaller than 
typical values of this variable (see middle and right panels of 
Fig.~3), and the probability for false detection 
is then extremely small.
Fig.~4 displays the probability distribution 
obtained with the three methods for the set of filtered light 
curves\footnote{The third method gives a probability which is always 
larger than $10^{-30}$. 
Indeed, the smallest value of $x$ which can be obtained
takes place when $\chi^2=0$, then $|x|\sim\sqrt{N_{m}}<10$.}. 

We apply the three algorithms which all give a probability for 
no periodicity. 
A star is accepted only if selected by all three methods with 
the following thresholds~: $P({\rm method \ 1}) < 10^{-15}$,
 $P({\rm method \ 2}) < 10^{-10}$, $P({\rm method \ 3}) < 10^{-5}$, 
tuned in order to allow one to retrieve the previously 
known Cepheids observed by {\sc EROS} in the Magellanic Clouds 
(see \citet{TheseDerue} for more details).
This procedure ({\it cut 4}) selects a sample of 2,553 stars.

Ten times more stars would have been selected if we had 
used method 1 or method 2 only (with the same thresholds), most of them being 
spurious variables.
\begin{figure}[h!]
\begin{center}
\resizebox{\hsize}{!}{\includegraphics{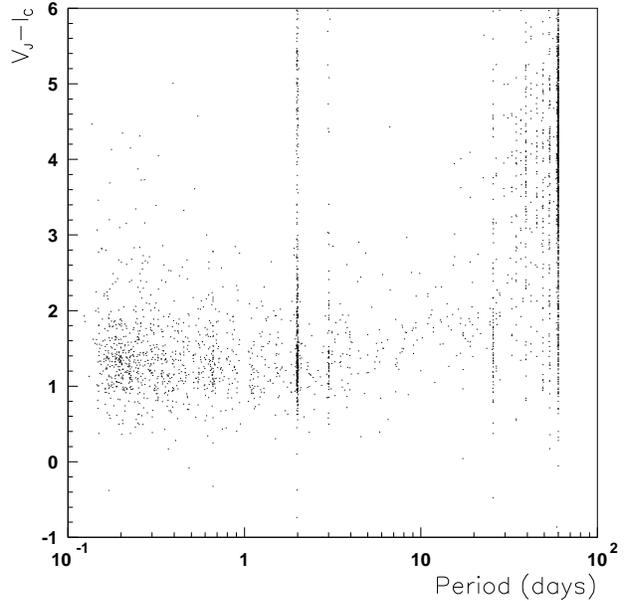}}
\caption[]
{Period-colour diagram ($P$ in days vs $V_{J}$-$I_{C}$) 
of the 2,553 selected candidates before cut 5.}
\end{center}
\label{aliasing}
\end{figure}
The third method would have added far less candidates if used alone, 
but still a factor 2 more.
Combining  independent methods has thus the advantage of 
considerably reducing the noise background 
(mostly due to aliases of one day or noisy measurements) while 
giving redundant information about the period.
To obtain individual periods we perform Fourier fits with 
five harmonics~:
\begin{equation}
{\rm flux} = {\rm flux}_{0} + 
\sum^{5}_{l=1} a_{l} \cos(\frac{2 \pi}{P} l (t-t_0) + \phi_{l}), 
\label{flux}
\end{equation}
where $P$ is the period, $\phi_{l}$ the phase and $a_{l}$ the amplitude. 
We define the amplitude ratios $R_{kl} = a_{k}/a_{l}$, 
and the phase differences (defined modulo 2$\pi$) 
$\Phi_{kl} = \phi_{k} - k \phi_{l}$, with $k>l$. 
Objects with non-significant harmonic amplitudes ({\it i.e} 
with almost sinusoidal light curves) have $R_{21} \sim 0$ and their 
$\Phi_{21}$ is ill defined.

The selection of periodic variable stars is complicated by aliases. 
Some of the stars with periods equal to a simple fraction or a low multiple 
of one day may be badly phased because of the nightly 
sequence of measurements.
These aliased periods are seen in Fig.~5 as vertical groups of dots 
at $2/3$, $2$ and $3$ days.
To eliminate them we demand ({\it cut 5}) 
that the fitted periods are not within 
$\pm 1\%$ of these values.
One can also notice some vertical groups of points around $25$ 
days which correspond to data gaps in our sample 
(see Fig.~1). 
Once these objects are removed, 2,424 stars remain.
The flux values of the remaining stars are folded using each period
obtained with the three methods.
The resulting phase diagrams are visually inspected. 
Some of them display an obvious spurious periodic or quasi-periodic
variability due to a low 
photometric quality.
After this final visual selection ({\it cut 6}) the list of variable 
stars includes 1,362 candidates which exhibit unambiguous periodic variability.
\subsection{Type of variability}
The classification of the selected stars among 
different types of variability cannot be based on the position of the objects 
in the colour-magnitude diagram since 
the spread in distance of these stars entails a spread in 
magnitude and colour.
It is desirable however to classify the various light curves according
to some physical parameters.
In the following we mainly 
use criteria based on the period $P$ of the luminosity variations 
and on the amplitude ratio $\Delta V_{E}/\Delta R_{E}$.
For each selected type of variable star the phase diagram of a typical 
candidate is displayed in Fig.~6.
\begin{figure*}
\begin{center}
\resizebox{\hsize}{!}{\includegraphics{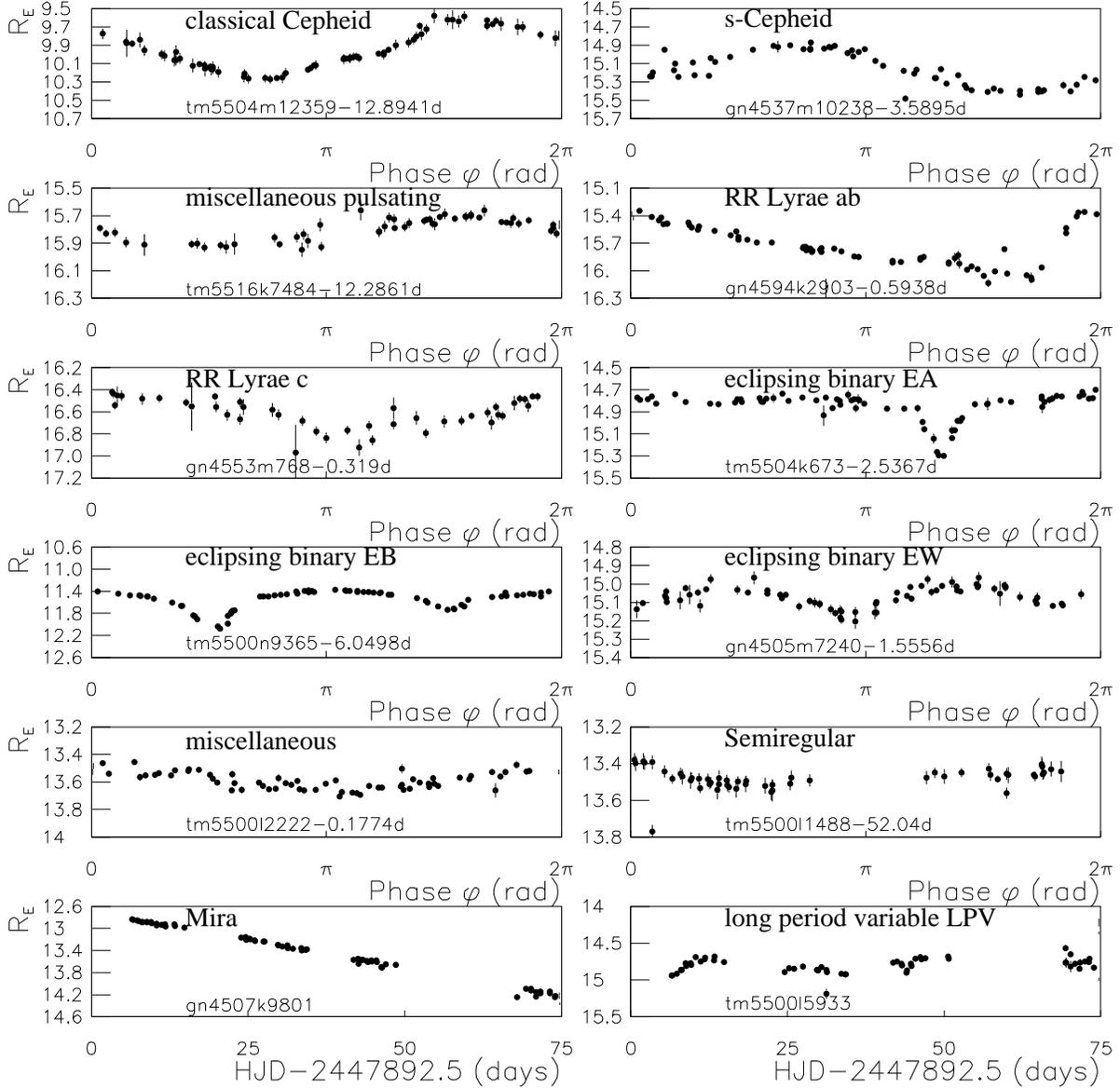}}
\caption[]{Phase diagrams (or light curves for the two lowest
panels) for typical variable stars of our catalogue 
in $R_{E}$ magnitude.}
\label{clum}
\end{center}
\end{figure*}
Three groups are distinguished depending on their period~:
\begin{itemize}
\item[$1^{\rm st}$~:] stars with a period larger than 60 days (543 objects)~:\\ 
For these objects an entire period has not been observed.
There is thus no warranty that these objects are periodic ones.
34 display a nearly linear light curve and are thus catalogued 
as Miras candidates.
The other 509 objects are catalogued as Long Period Variable stars (LPVs).
\item[$2^{\rm nd}$~:] stars with a period between 30 and 60 days (387 objects)~:\\
264 Semi-Regular variable stars are selected by requiring~:
$\Delta V_{E}$/$\Delta R_{E} >$ 1.2 to select pulsating stars (see below),
$V_{J}$-$I_{C}$ $>$ 2.5 to discriminate from bluer variable stars and 
$\Delta R_{E} <$ 1.0 to avoid possible Miras or LPVs wrongly phased. 
The long term stability of these stars is not known. 
Some of the reported periods may change from season to season, 
as a result of their semi-regular behaviour.
The remaining 123 objects are catalogued as miscellaneous 
variable stars. 
%
\begin{table}[h!]
\caption[]
{Number of selected objects for each type of variability.}
\label{tabvariables}
\begin{flushleft}
\begin{center}
\begin{tabular}{lll}
\hline
Period range & Type &         Number of objects                       \\ \hline
$P>60$ d       &                    & 543 \\ \hline
               & LPV                & 509 \\
               & Miras              &  34 \\ \hline	       
$60 > P>30$ d  &                    & 387 \\ \hline
               & Semi-Regular       & 264 \\ 
	       & miscellaneous      & 123 \\ \hline
$P<30$ d       &                    & 432 \\ \hline
	       & pulsating          &  60 \\ \hline   
               & RR$c$              &  14 \\
               & RR$ab$             &   5 \\
               & classical-Cepheids &   6 \\
               & $s$-Cepheids       &   3 \\
               & miscellaneous      &  32 \\ \hline
	       & non-pulsating      & 372 \\ \hline
               & EA                 & 130 \\
               & EB                 &  35 \\
               & EW                 &  11 \\
               & miscellaneous      & 196 \\
\hline
\end{tabular}
\end{center}
\end{flushleft}
\end{table}
\item[$3^{\rm rd}$~:] stars with a period smaller than 30 days (432 objects)~: \\
The colour change for a Cepheid in standard passbands 
is $\Delta V_J$/$\Delta I_{C}>$1.3 \citep{madore91} which corresponds 
to $\Delta V_{E}$/$\Delta R_{E}>$ 1.2 in the {\sc EROS} system. 
Two sets are thus distinguished based on this criterion~: 
\begin{itemize}
\item[-] The pulsating variable stars (60 objects)~:\\
For stars with period $P<1$ day, two samples of RR Lyr{\ae} are identified~:
the RR$c$ have $ R_{21} < 0.4$ (14 objects) and 
the RR$ab$ have $R_{21} > 0.4$ (5 objects).
\begin{figure}[h!]
\begin{center}
\resizebox{\hsize}{!}{\includegraphics{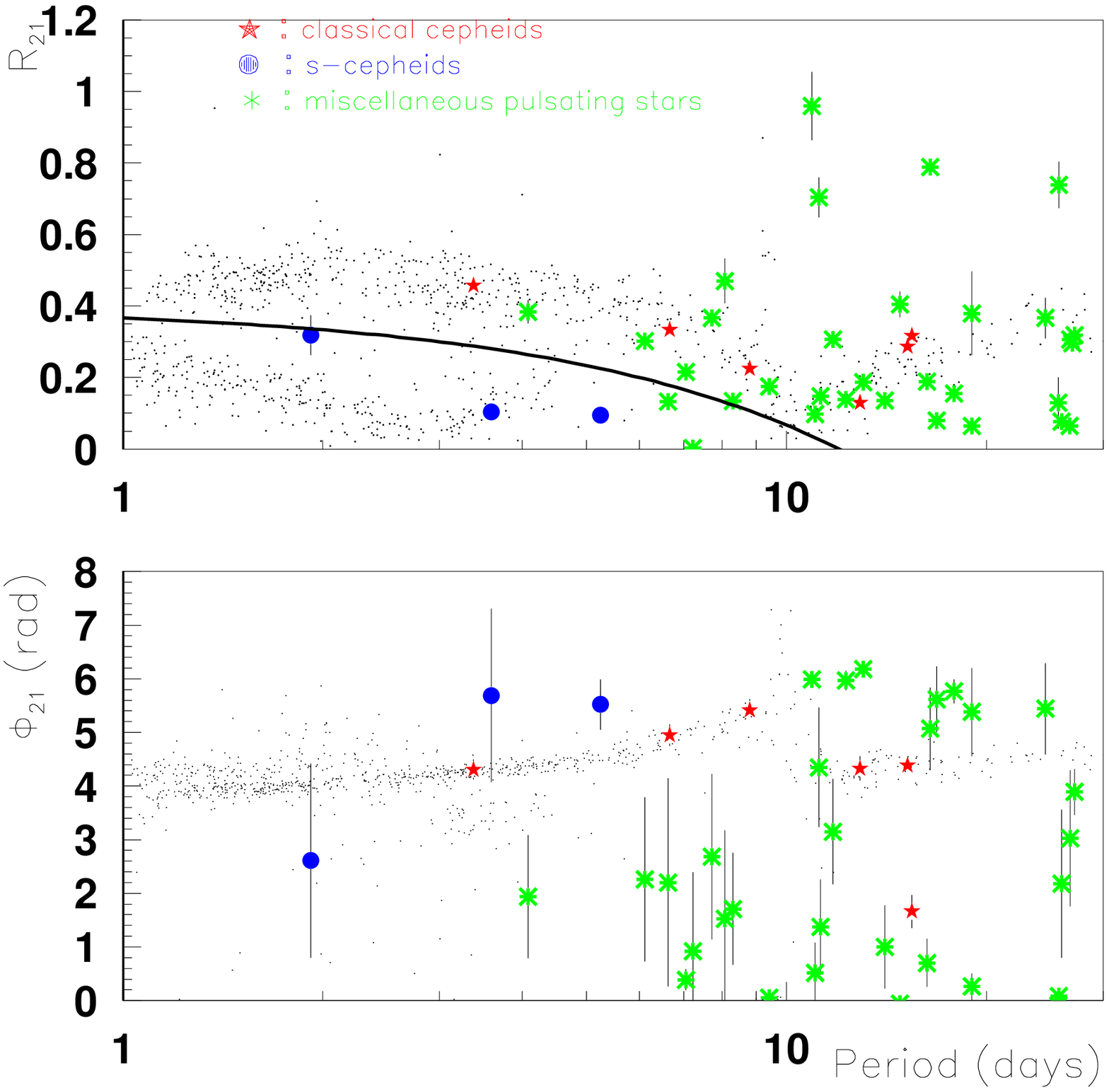}}
\caption[]{Distinction between classical Cepheids, 
$s$-Cepheids and miscellaneous 
pulsating stars in the 
$R_{21}-P$ and $\Phi_{21}-P$ planes. 
The individual uncertainties are reported.
The cloud of points represent the Cepheids observed in the {\sc LMC} 
(adapted from \citet{afonso-1999}).
The curve corresponds to the empirical 
function $R_{21}^{\rm cut}(P)=0.4 - (P/30$ days) used to distinguish 
between $s$ and classical Cepheids. 
}
\end{center}
\label{r21-p}
\end{figure}
We adopt the morphological classification 
proposed by \citet{Antonello} and classify as $s$-Cepheids the stars 
that lie in the lower part of the $R_{21}-P$ plane, and as classical 
Cepheids the remaining stars. 
$s$-Cepheids pulsate in the first overtone and classical Cepheids in the 
fundamental mode (see {\it e.g} \citet{beaulieu1995, beaulieu1996a}).
We use the empirical function $R_{21}^{\rm cut}(P)=0.4 - (P/30$ days) 
in order to separate these pulsation modes (see Fig.~7). 
Among the five objects that pass the $s$-Cepheid cut, only three 
belong to the $R_{21}-P$ and $\Phi_{21}-P$ 
distributions of galactic $s$-Cepheids and their phase parameter 
$\Phi_{21}$ is poorly constrained.
\begin{figure*}[h!]
\begin{center}
\resizebox{\hsize}{!}{\includegraphics{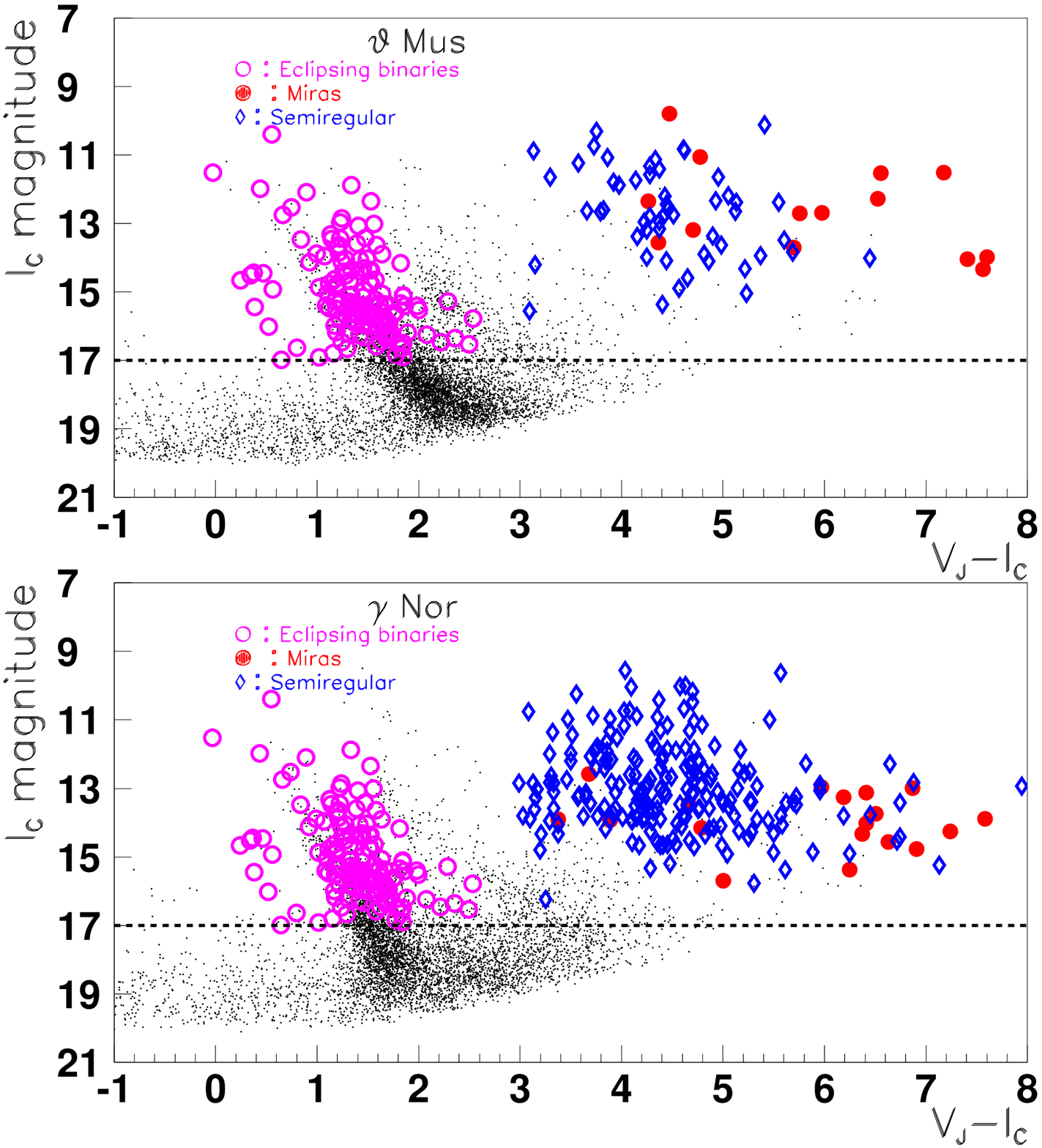}}
\caption[]{Colour-magnitude diagrams ($I_{C}$ vs $V_{J}$-$I_{C}$) for the 
Miras (represented by dots $\bullet$), Semi-Regular 
variable stars (diamonds $\diamond$) and 
eclipsing binaries (open circles $\circ$).
The dotted line corresponds to cut 2 on the luminosity 
of the stars.
}
\end{center}
\label{diagHR1}
\end{figure*}
\begin{figure*}[h!]
\begin{center}
\resizebox{\hsize}{!}{\includegraphics{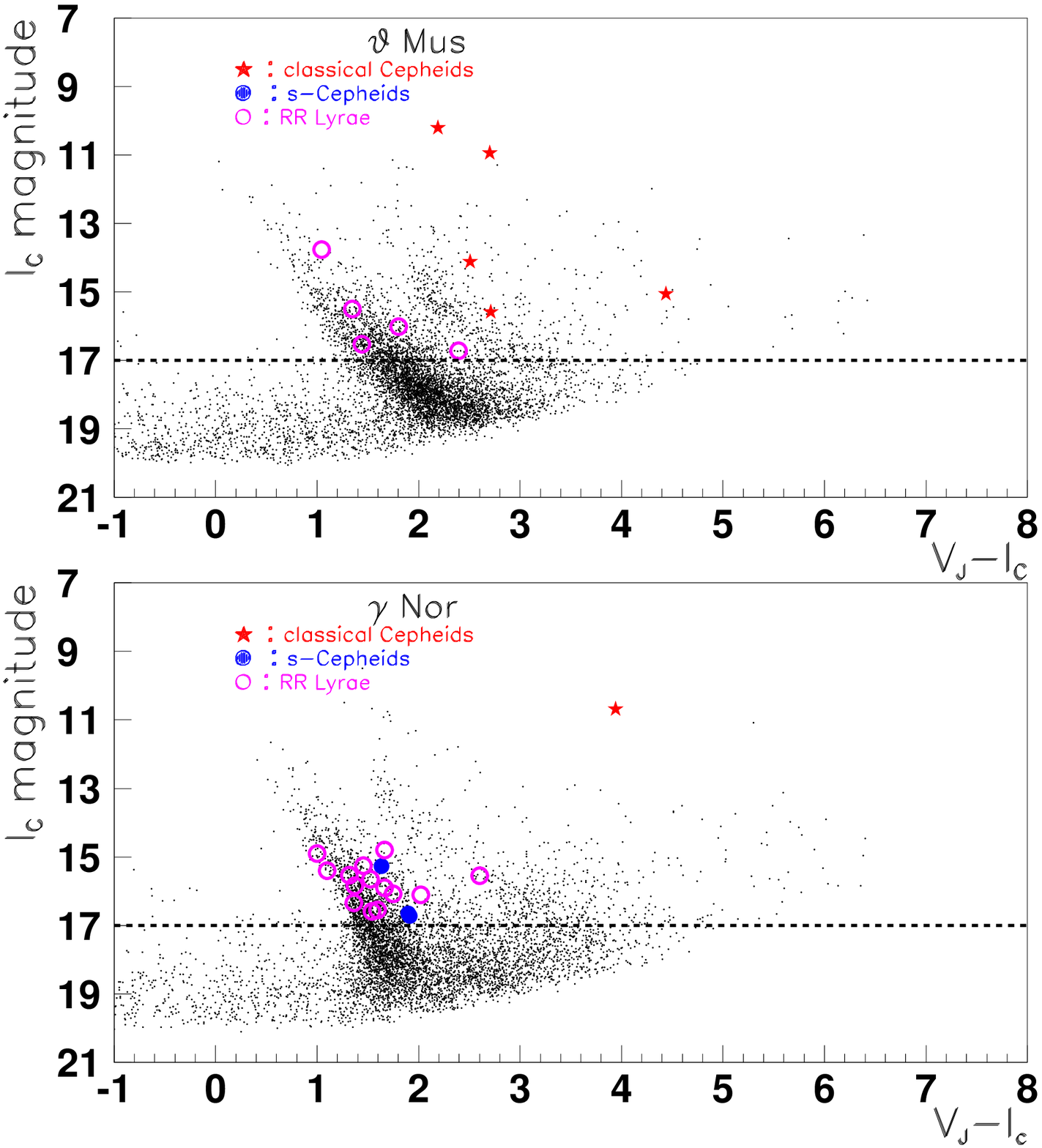}}
\caption[]{Colour-magnitude diagrams ($I_{C}$ vs $V_{J}$-$I_{C}$) for the 
classical Cepheids (represented by stars $\star$), 
$s$-Cepheids (dots $\bullet$), 
and RR Lyr{\ae} (open circles $\circ$).
}
\end{center}
\label{diagHR2}
\end{figure*}
Besides most of classical Cepheids have amplitudes larger than 
$\Delta R_E >0.4^{\rm mag}$ (see {\it e.g} \citet{afonso-1999}).
For stars with period $P>1$ day, three samples are then identified~:
the classical Cepheids have $R_{21} > R_{21}^{\rm cut}$ 
and $\Delta R_{E} > 0.4^{\rm mag}$ (6 objects);
the $s$-Cepheids have $R_{21} < R_{21}^{\rm cut}$ 
and $3 \ {\rm rad} < \Phi_{21} < 6 \ {\rm rad}$ (3 objects);
the remaining 32 objects are catalogued as miscellaneous 
pulsating stars.
\item[-] The non-pulsating variable stars (372 objects)~:\\
The remaining objects have similar amplitudes in both passbands.
We classify them according to the following criteria. 
Algol systems (130 stars, type $EA$) display well-defined eclipses 
whose secondary one has a depth lower than half the primary one, 
and possibly flat light curve between them.
$EB$ type objects (35 in total) show a secondary 
eclipse equal to half the primary one. 
The $EW$ type (11 objects) is characterised by similar depths 
of the two eclipses. 
The members of a residual sample of 196 objects do not look 
like convincing 
eclipsing binaries and are catalogued as miscellaneous variable stars.
\end{itemize}
As emphasised by \citet{Udalski-1999a} a large number of 
variable objects show small amplitude sinusoidal variations, such as 
ellipsoidal binary variable stars. 
A contamination of the sample of pulsating stars by eclipsing binaries 
is thus possible.
\end{itemize}
Figures 8 and 9 show the location of the selected variables in the 
colour-magnitude diagrams.
Also plotted are 10,000 stars located 
in the central part of the two fields tm550 and gn450.
Most of the Cepheids are much brighter than our magnitude threshold 
(cut 2); this is not so for RR Lyr{\ae} (see Fig. 9, lower panel).
Our catalogue is thus not complete for this
type of variable stars, as already mentionned.
\section{The catalogue \label{catalog}}
The catalogue is composed of two tables containing 
objects with periods $P$ smaller or larger than 30 days, respectively.
The identifier of each star is given according to the recommendations of 
the {\sc IAU} Commission 5 in {\it The Rules and Regulations for 
Nomenclature} (see the Annual Index of A\&A). 
The general acronym used in the 
catalogue is {\sc EROS2 GSA} followed by J2000 equatorial coordinates 
in the format JHHMMSS$\pm$DDMMSS. 
The remainder of the identifier in parentheses gives some information 
relating to the internal organisation of the {\sc EROS} database~: 
gn$nnn$ or tm$nnn$ is the name of the field, followed by the CCD number and 
the location on the image following the {\sc EROS II} 
nomenclature.
The remaining number is the star identifier used in the {\sc EROS} database. 
As an example, J132630-630945(tm5504m12359) is the 
name of the 12359th star 
observed in quarter {\rm m} of CCD 4 in the field tm550. 
The J2000 equatorial coordinates of this star are 13:26:30.11, -63:09:45.66.

The equatorial coordinates (J2000) of individual stars have been 
obtained as follows. 
First, we have inserted the suitable {\sc WCS} keywords into the header of 
the {\sc EROS II} reference images using the {\sc WCSTools} 
package \citep{wcstools}.
Whenever possible, the cross-identification of each star with previously known 
objects within a 10\arcsec \ search radius has been done using 
the {\sc Simbad} and {\sc Vizier} databases available at 
the {\sc CDS}, Strasbourg.

The tables contain the following information~:
\begin{enumerate}
\item[1.] Identifier
\item[2.] Right ascension $\alpha$ (J2000)
\item[3.] Declination $\delta$ (J2000)
\item[4.] $\langle R_{E} \rangle $ mean magnitude in EROS-red
\item[5.] $\Delta R_{E}$ amplitude peak to peak in $R_{E}$
\item[6.] $\langle V_{E} \rangle $ mean magnitude in EROS-visible
\item[7.] $\Delta V_{E}$ amplitude peak to peak in $V_{E}$
\item[8.] Period in days. Note that periods longer than 
30 days are given with less accuracy since the time span 
of the measurements does not allow a precise determination. 
Measured periods which are longer than 60 days 
({\it i.e} 2/3 of the observation 
period) are flagged by writing ``$P>60$ d'' and have no warranty to 
be true periodic variable stars.
The peak to peak amplitude of these stars could 
be meaningless and the mean magnitude is determined with low accuracy.
\end{enumerate}
For Cepheids and RR Lyr{\ae} the results of the Fourier fit are given~:
\begin{enumerate}
\item[9.] Fourier coefficient ratio $R_{21}$;
\item[10.] Fourier coefficient ratio $R_{31}$;
\item[11.] Phase difference $\phi_{21}$ (in {\rm rad});
\end{enumerate}
Also given when possible~:
\begin{enumerate}
\item[13.] Type of variability (C=classical Cepheids, 
S=$s$-Cepheids, puls.= miscellaneous pulsating stars, 
EA,EB,EW=eclipsing binaries, misc=miscellaneous 
variable stars, SR=Semi-Regular variables, M=Miras, 
LPV=long period variables);
\item[14.] Name of cross-identified object(s) 
within a search radius of 10\arcsec .
\end{enumerate}
The catalogue is planned to be installed 
at the {\sc CDS} 
(see also our Web site {\tt http://eros.in2p3.fr/}).

A by-product of such a survey 
is the possibility to update the coordinates given in older catalogues.
As an example, it was found that the well known 
classical Cepheid OO Cen with the {\sc SIMBAD} 
identifier V*OO Cen was 20" away from an {\sc EROS}
object, namely J132630-630945(tm5504m12359), as shown in
the finding chart of Fig.~10.
%
\begin{table}[h!]
\caption[]
{Period and coordinates obtained on OO Cen by {\sc SIMBAD}, {\sc GCVS} and this study.}
\label{taboocen}
\begin{flushleft}
\begin{center}
\begin{tabular}{llll}
\hline
Source           & {\sc SIMBAD} & {\sc GCVS} & This study \\ \hline
object           & V* OO Cen    & OO Cen     & J132630-630945(tm5504m12359) \\
P (days)         & n/a          & 12.8805    & 12.894 \\
$\alpha$ (B1950) & 13:23:06.85  & 13:23:09   & 13:23:08.8 \\ 
$\delta$ (B1950) & -62:54:24.9  & -62:54.0   & -62:54:11 \\ \hline
\end{tabular}
\end{center}
\end{flushleft}
\end{table}
\begin{figure}[h!]
\begin{center}
\resizebox{\hsize}{!}{\includegraphics{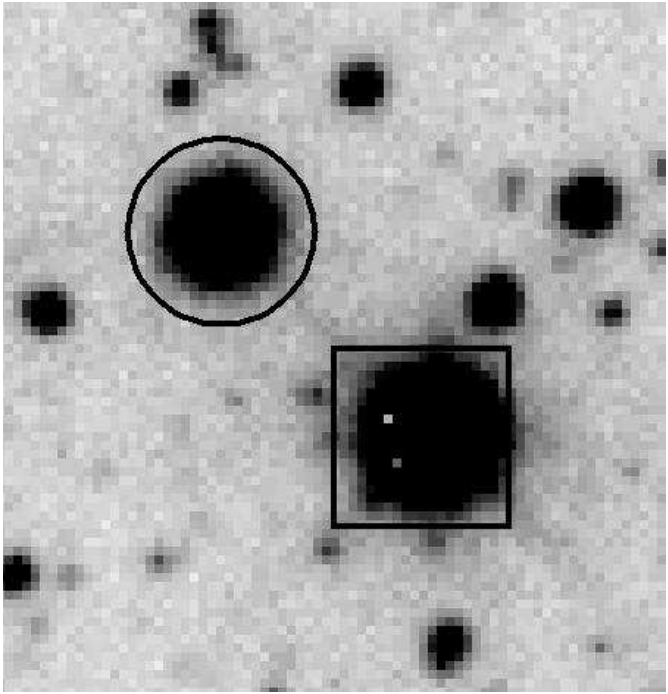}}
\caption[]{Finding chart of the star
OO Cen - J132630-630945(tm5504m12359).
The position given by the {\sc SIMBAD} database is shown by the square
which is 10" wide, while the {\sc EROS} position is
shown by the circle. 
}
\end{center}
\label{findingchart}
\end{figure}
The period and coordinates we report are compatible with 
the ones given by the {\sc General Catalogue 
of Variable Stars (GCVS)}\footnote{Queries on this catalogue are possible 
at http://www.sai.msu.su/cgi-bin/wdb-p95/gcvs/stars/form} 
\citep{Kholopov-1985} (see Table \ref{taboocen}).
It seems that an error occurred when the coordinates of this particular 
star were filled in the {\sc SIMBAD}. 
The light curve of this object is shown in Fig~6.

We have performed several cross-identifications of our 
catalogue with those previously available, namely 
the {\sc IRAS Point Source Catalog} \citep{beichman},
the {\sc MSX5C Infrared Astrometric Catalog} \citep{egan99}, 
the {\sc Two Micron All Sky Survey (2MASS)} \citep{2mass},
the {\sc CKS91} catalogue \citep{cald91} and the 
{\sc General Catalogue of Variable Stars (GCVS)} \citep{Kholopov-1985}.
A total of 38 {\sc IRAS} sources and 220 {\sc MSX5C} objects 
have been thus retrieved. 
The overlap with the available {\sc 2MASS} catalogue exists only for the 
{\sc EROS} field gn459, representing 255 stars. 
A total of 233 {\sc 2MASS} objects have been thus retrieved among them
37 objects classified as Semi-Regular variables.

An overlap exists with the {\sc CKS91} catalogue \citep{cald91}. 
These authors have searched for bright Cepheids and 
other variable stars with 
$I < 14$, towards Crux and Centaurus, during 42 days with less than 10 
measurements per star.
A small overlap exists between this survey and the GSA fields 
tm550 and tm551.
Unfortunately this overlap involves our CCD \#2 which was not operational 
at the time of the observations. 
Therefore the comparison can only be carried out on $0.7$ square degree. 
Furthermore, as pointed above, this comparison is restricted to 
stars with magnitude $I_C < 14$.
A total of 118 {\sc EROS} objects, 6 {\sc GCVS} and 23 {\sc CKS91} 
objects lie in this region.
Three objects are common to the {\sc GCVS} and 
{\sc CKS91} catalogues one.
The overlap between the three catalogues represents only 19 objects.
Among them one finds the most interesting 
ones, such as Cepheids {\sc OO Cen} and {\sc V881 Cen} 
(see Table \ref{varretrouve}).
Some objects show a large difference between the 
two magnitude determinations. 
These are long period variables 
for which the mean magnitude is measured on only a 
part of the whole period, and thus ill determined in both surveys.
Only 7 known variable stars are not recovered by our analysis
(see Table \ref{varperdu}). 
Conversely, 99 objects of our catalogue are not listed 
by {\sc CKS91}. All of them are labelled Long Period Variable stars (LPV).
{\scriptsize
\begin{table*}[H!]
\begin{center}
\caption{Previously known variable stars recovered by this survey. 
For each star we give its name, the original catalogue, the type 
of variability, the period and the $I$ magnitude given (if any) 
in the catalogue and in this survey.}
\label{varretrouve}
\begin{tabular}{|l|c|c|c|c|c|c|c|}
\hline
Object             & Catalogue & Type & P(days)  & I  & EROS ID   & P(days) & $I_c$ \\
\hline
OO Cen           & GCVS   & C    & 12.8805 & 9.96  & J132630-630945 & 12.8941 & 10.20 \\
           &    &     &  &   & (tm5504m12359) &  &  \\
V881 Cen         & GCVS   & C    &    -    &10.57  & J132721-630110 & 15.2278 & 10.97 \\
         &    &     &       &  & (tm5503l3554) &  &  \\
V608 Cen         & GCVS   & EB   & 1.6287  & 12.95 & J132948-630634 & 1.7601 & 12.05 \\
        &   &   &  & & (tm5505m13922) &  &  \\
CKS91  & CKS91 & EA   &       -        & 13.29 & J132758-631449 & 5.4918 & 13.03 \\
13246-6259 &  &    &               &  & (tm5505l9598) &  & \\
CKS91  & CKS91 & E    &       -        & 13.05 & J133311-630023 & 0.2141 & 13.20 \\
13297-6244 &  &     &               & 1 & (tm5511m5801) & &  \\
CKS91  & CKS91 & SR   &       -        & 12.35 & J132400-632057 & 45.83  & 12.34 \\
13206-6305 &  &   &              &  & (tm5504l1003) &   &  \\
CKS91  & CKS91 & LPV  &       -        & 11.73 & J132427-631542 & $P>60$ & 11.75 \\
13211-6300 &  &   &              &  & (tm5504k5591) &  &  \\
CKS91  & CKS91 & SR   &       -        & 10.51 & J132444-632224 & 55.00  & 10.79 \\
13214-6306 &  &   &               &  & (tm5504l7786) &   &  \\
CKS91  & CKS91 & SR   &       -        & 12.85 & J132447-631201 & 52.39  & 12.77 \\
13214-6256 &  &    &              &  & (tm5504k8841) &   &  \\
CKS91  & CKS91 & LPV  &       -        & 10.43 & J132454-631437 & $P>60$ & 10.23 \\
13215-625 & &  &             &  &(tm5504k9938)  &  &  \\
CKS91  & CKS91 & LPV  &       -        & 11.81 & J132519-631720 & $P>60$ & 13.00 \\
13219-6301 &  &  &              &  & (tm5504l13138) &  &  \\
CKS91 & CKS91 & LPV  &       -        & 12.51 &  J132433-632201 & 58.46 & 11.96 \\
13229-6252 &  &   &               &  & (tm5504m10330)  &  &  \\
CKS91  & CKS91 & LPV  &       -        & 12.48 & J132625-630645 & $P>60$ & 12.41 \\
13230-6251 &  &   &              &  & (tm5504m11603) &  &  \\
CKS91  & CKS91 & LPV  &       -        & 12.63 & J132714-634338 & $P>60$ & 10.78 \\
13238-6227 &  &   &               &  & (tm5507l1058) &  &  \\
CKS91  & CKS91 & LPV  &       -        & 12.21 & J132719-633644 & $P>60$ & 11.91 \\
13238-6254 &  &   &               &  & (tm5507l1752) &  &  \\
CKS91  & CKS91 & M    &       -        & 12.46 & J132803-632200  & $P>60$ & 13.99 \\
13246-6306 & &     &         &  & (tm5505l10106)  &  &  \\
CKS91  & CKS91 & SR    &       -       & 12.66 & J132813-630013 & 44.80 & 12.72 \\
13247-6244 &  &    &         &  & J132813-630013 &  &  \\
CKS91  & CKS91 & LPV  &       -        & 12.37 & J132948-631227 & $P>60$ & 12.56 \\
13264-6256 &  &   &      &  & (tm5510n4312) &  &  \\
CKS91  & CKS91 & M    &       -        & 12.65 & J133222-632339 & $P>60$ & 12.65 \\
13289-6308 & &   & &  & (tm5513k12129) &  &  \\
\hline  
\end{tabular}
\end{center}
\end{table*}
}
\begin{table}[h!]
\begin{center}
\caption{Previously known variable stars {\it not} recovered by this survey. 
For each of them we give (where possible) their type of variability, 
their period and the reason for their absence in the catalogue.}
\label{varperdu}
\begin{tabular}{|l|c|c|c|l|}
\hline
Object     & Type    & P(days) & catalogue & comment\\
\hline
HQ Nor     & EB      & 90.9   & GCVS  & too long period   \\
HY Nor     & Mira    & 236    & GCVS  & close to \\
           &         &        &       & the CCDs gap   \\
UW Nor     & EA      & 8.4860 & GCVS  & fails cut 6\\
13214-6256 &  LPV    &        & CKS91 & fails cut 6 \\
13218-6254 &  LPV    &        & CKS91 & fails cut 6 \\
13248-6249 &  LPV    &        & CKS91 & fails cut 6 \\
13232-6249 &  LPV    &        & CKS91 & fails cut 6 \\
\hline  
\end{tabular}
\end{center}
\end{table}
\clearpage
\section{Discussion}\label{discussion}
The motivation of our search was to improve our knowledge 
of the distance distribution and of the extinction of the 
microlensing source stars used in papers I and II.
In the following we use the RR Lyr{\ae} which are 
well-known distance indicators and have been 
observed in all six directions that we investigated.
The GSA fields having a high non-uniform absorption, we give only 
an average reddening towards our fields, based on {\sc EROS} data alone.

For each selected RR Lyr{\ae} we estimate the extinction 
$A(V) = 2.4 \times E(V-I)$
using the standard extinction coefficients \citep{schleg98, stanek-1996}.
The colour excess $E(V-I)$ is derived from the colour that 
we measured and their intrinsic colour~:
$\left( V - I \right)^{RRab}_{0} = 0.4$ and
$\left( V - I \right)^{RRc}_{0} = 0.2$ \citep{alcock-1998}.
The uncertainty on individual extinctions
is estimated to be $0.5^{\rm mag}$. 
This error includes the $0.2^{\rm mag}$ uncertainty on the 
{\sc EROS} colour measurement (see Eq.~(\ref{colour}))
and a $0.1^{\rm mag}$ magnitude uncertainty on the intrinsic colour.
The absolute magnitude of RR Lyr{\ae} is 
$M_{V}^{RR} \sim 0.71$
with a precision of $0.07^{\rm mag}$ \citep{gould-1998}.
Finally, we estimate the distance to each star simply 
by using the relation $m - M = 5 \times \left( \log(d)-1 \right) + A$.
The typical uncertainty on the distance is 20\%. 
The left panels of Fig.~11 show the obtained extinction 
$A(V)$ versus the calculated distance. 
The mean extinction of the RR Lyr{\ae} 
is $A(V) \sim 3.6 \pm 0.2$ magnitude towards $\theta$ Mus and 
$3.3 \pm 0.1$ towards $\gamma$ Nor, with a dispersion of 1 magnitude.
The mean distance of the RR Lyr{\ae} is $\sim$4.7$\pm$ 0.3 {\rm kpc} 
towards $\theta$ Mus and $5.3 \pm 0.2$ {\rm kpc} towards $\gamma$ Nor.
The dispersion of the values is 1.4 {\rm kpc} which reflects the spread in 
distance of disc stars.
The disc population contributes to the sources of the microlensing events
that we observe in the Galactic plane. 
A model based on the distribution of matter in the disc 
and the luminosity function of neighbooring stars has been used 
in \citet{TheseDerue} to estimate the distance of the star 
population from the disc. 
The distance of RR Lyr{\ae} is in good agreement with the one obtained 
with this model.
\begin{figure}
\begin{center}
\resizebox{\hsize}{!}{\includegraphics{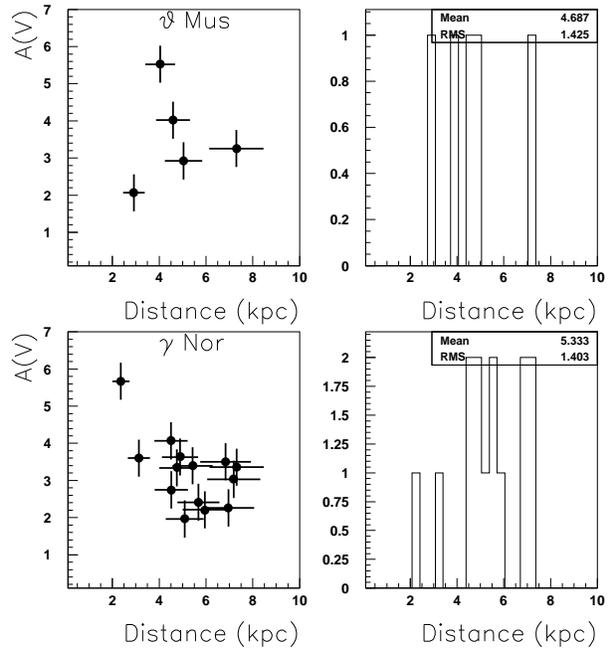}}
\caption[]{Extinction $A(V)$ versus the distance of RR Lyr{\ae}(in {\rm kpc}) 
towards $\theta$ Mus and $\gamma$ Nor. 
The right panels show the inferred distance distribution.
}
\end{center}
\label{extinction}
\end{figure}
\section{Conclusion}
In the course of our program dedicated to microlensing events, 
we have devoted a fraction of observing time to the search for variable 
stars in six directions of the Galactic plane. 
This exploratory campaign, that lasted three months, 
led to the discovery of 1,362 variable stars. 
Among them we identified 9 Cepheids, 19 RR Lyr{\ae}, 34 Miras,
176 eclipsing binaries and 266 Semi-Regular variable stars.
We have set up a catalogue of all of the 1,362 stars and
cross-identified it with several other catalogues. 
In particular a comparison with the {\sc GCVS} and 
the {\sc CKS91} catalogues
shows that only a small fraction ($\sim$15\%) of the objects that we
have identified appear in those two.
Among the stars most appropriate to be used as distance indicators,
the Cepheids turned out to be too few to warrant a
particular study. 
As far as RR Lyr{\ae} are concerned, we have
determined their mean distance and found it to be $\sim$5 {\rm kpc}.
Yet the statistics being quite limited, we are considering pursuing 
this effort by launching a longer search for variable stars based 
on the results of this first campaign.
\begin{acknowledgements}
The WCSTools package was made available to us thanks to the work of Doug Mink, 
{\sc NASA/GSFC}, Harvard. 
The Skycat/Gaia tool is the result of a joint effort 
by the computer staff of {\sc ESO} Garching and from the Starlink Project, UK.
\end{acknowledgements}

\end{document}